\documentclass[prd,superscriptaddress,amsfonts,amssymb,amsmath,showpacs,twocolumn]{revtex4-2}
\usepackage{bm}
\usepackage{amsfonts}
\usepackage{latexsym}
\usepackage{graphicx}
\usepackage{amsmath}
\usepackage{palatino}
\usepackage{mathpazo}
\usepackage{textcomp}
\usepackage{float}
\usepackage{booktabs}
\usepackage{dcolumn}
\usepackage{booktabs}
\usepackage{multirow}
\usepackage{hyperref}
\usepackage{orcidlink}
\hypersetup{colorlinks,citecolor=blue}
\usepackage{xcolor}

\begin{document}

\color{black}       

\title{Probing Dark Energy Properties in $f(Q,C)$ Gravity with FLRW Cosmological Models}

\author{ N. Myrzakulov\orcidlink{0000-0001-8691-9939}}\email{nmyrzakulov@gmail.com}
\affiliation{L N Gumilyov Eurasian National University, Astana 010008, Kazakhstan.}
\author{Anirudh Pradhan\orcidlink{0000-0002-1932-8431}}
\email{pradhan.anirudh@gmail.com}
\affiliation{Centre for Cosmology, Astrophysics and Space Science (CCASS), GLA University, Mathura-281406, U.P., India.}
\author{A. Dixit\orcidlink{0000-0003-4185-4162}}
\email{archana.ibs.maths@gmail.com}
\affiliation{Department of Mathematics, Gurugram University Gurugram, Haryana-122003, India.}
\author{ S. H. Shekh\orcidlink{0000-0002-1932-8431}}\email{  da\_salim@rediff.com}
\affiliation{L N Gumilyov Eurasian National University, Astana 010008, Kazakhstan}
\affiliation{Department of Mathematics, S.P.M. Science and Gilani Arts, Commerce College, Ghatanji, Yavatmal, \\Maharashtra-445301, India.}

\begin{abstract}
\textbf{Abstract:} This study delves into the cosmological implications of the $f(Q,C)$ modified gravity framework within the context of the FLRW spacetime which offers a dynamic alternative to the standard $\Lambda$CDM cosmology. Here, we define the transit form of Hubble's parameter to explain several geometrical and physical aspects. The chosen parametric form of the Hubble parameter represents a smooth transition from the decelerating early universe to the accelerating present and late-time evolution. Employing observational datasets such as the Hubble parameter, Type Ia supernovae, Baryon Acoustic Oscillations (BAO), and Standard Candles (SC), we constrain the model parameters using the Markov Chain Monte Carlo (MCMC) method. The isotropic pressure, energy density, equation of state parameter, and energy conditions were analyzed to explore the physical viability of the $f(Q,C)$ framework. The results highlight the model's ability to replicate key cosmological behaviors, including the accelerated expansion driven by dark energy. 
\newline
\textbf{Keywords:} FRW universe; $f(Q,C)$ gravity; cosmology; dark energy
\end{abstract}

\maketitle

\section{Introduction}
In the last ten years, cosmological theories have suggested that Dark Energy (DE) dominates the current universe, causing an accelerated expansion \cite{1,2,3,4,5,6,7,8,9,10,11}. Recent observations in cosmology indicate that the universe is spatially flat, with an estimated composition of approximately 70\% DE, 30\% matter—primarily Cold Dark Matter (CDM) and baryons—and an insignificant amount of radiation. While it's clear that DE plays a key role in the universe's fate, its exact nature and origins remain unknown. Several models have been proposed to describe or explain DE, including various hypotheses in the literature. Some of these include the evolving canonical scalar field, known as quintessence, with an Equation of State (EoS) parameter within the range $-1<\omega <-\frac{1}{3}$; phantom energy, which has an EoS parameter less than $\omega <-1$ and violates the weak energy condition (WEC); and quintom energy, where the EoS evolves through $\omega = -1$ \cite{12,13,14,14a,14b,15, 16,17,18}. The current EoS parameter for DE has been estimated from combined data from WMAP9 (the Nine-Year Wilkinson Microwave Anisotropy Probe), along with $H_0$ measurements, Type Ia Supernova (SNIa), the Cosmic Microwave Background (CMB), and Baryon Acoustic Oscillations (BAO). These observations suggest a value of $\omega_0 = -1.084 \pm 0.063$ \cite{7}. The Planck collaboration in 2015 updated this estimate to $\omega_0 = -1.006 \pm 0.0451$ \cite{19}, with further refinement in 2018 showing $\omega_0 = -1.028 \pm 0.032$ \cite{19}.\\
In recent years, researchers have introduced several alternative approaches to the standard Einstein-Hilbert action, which forms the basis of general relativity, to address the issue of the universe's accelerating expansion. These approaches are collectively referred to as modified theories of gravity (MTG). Numerous distinct actions have been put forward under MTG to provide explanations for this cosmic acceleration. Some of the most commonly explored MTG are: \textbf{\textit{$f(R)$ Gravity}}: $f(R)$ gravity generalizes general relativity by replacing the Ricci scalar $R$ in the Einstein-Hilbert action with a general function $f(R)$. This is done to explain cosmic acceleration without dark energy \cite{20,21,21a,21b,21c,21d,21e,21f,21g}. \textbf{\textit{$f(T)$ Gravity}}: $f(T)$  gravity modifies the teleparallel equivalent of general relativity (TEGR) by replacing the torsion scalar $T$ with a general function $f(T)$. In this theory, torsion, rather than curvature, explains gravity \cite{22,23,24,25,26,27,28,29,29a,29b,29c}. \textbf{\textit{$f(G)$ Gravity}}: $f(G)$  gravity, the action is modified by introducing a general function of the Gauss-Bonnet scalar $G$, which combines the Ricci scalar and the Riemann curvature tensor \cite{30,31,32,33,34,34a,34b,34c,34d,34e}. \textbf{\textit{$f(Q)$ Gravity}}: $f(Q)$  gravity uses the non-metricity scalar $Q$ to describe gravitational interactions. The theory is part of the metric-affine geometry framework, where gravity is tied to changes in vector lengths rather than their directions \cite{35,36,37,38,39,40,41,41a,41b,41c,41d,41e,41f,41g,41h}. \textbf{\textit{$f(Q,T)$ Gravity}}: $f(Q,T)$ gravity extends $f(Q)$ gravity by including the trace of the energy-momentum tensor $T$ in the action. This allows for coupling between matter and geometry \cite{42,43,44,45,46}. Next, The \textbf{\textit{$f(Q,C)$ gravity}} / theory has been proposed to provide new insights into the nature of dark energy and the accelerating expansion of the universe. By extending beyond the linear dependence on $Q$, it opens up possibilities for explaining late-time cosmic acceleration without invoking exotic matter fields or a cosmological constant. The inclusion of $C$ also allows for new gravitational interactions that can be tested against observational data, such as those from cosmic microwave background radiation, large-scale structure, and type Ia supernovae. The non-metricity scalar $Q$ measures the deviation of the metric from being preserved under parallel transport in a given connection. Unlike General Relativity, where the connection is symmetric and torsion-free and the boundary term $C$ arises from the interplay between the torsion-free, curvature-free connection and the total divergence of certain quantities. It ensures that the theory is dynamically equivalent to General Relativity in special cases, allowing for a smooth transition between different geometric interpretations of gravity. The term $C$  can provide additional degrees of freedom, influencing the behavior of gravitational fields, particularly in cosmological contexts. One of the key motivations for $f(Q,C)$ gravity is the unification of different geometric formulations of gravity-curvature-based, torsion-based, and non-metricity-based theories. By including both the non-metricity scalar $Q$ and the boundary term $C$, the theory creates a framework that can interpolate between teleparallel gravity, general relativity, and other modified gravity theories. This makes $f(Q,C)$ a promising candidate for exploring the underlying geometric structure of spacetime.\\
The gravitational action of the $f(Q, C)$ gravity theory is given by

\begin{equation}\label{1}
	S=\; \int \bigg(\frac{1}{2k}f(Q,C)+\mathcal{L}_{m}\bigg)\sqrt{-g}d^{4}x,
\end{equation}
The field equation can be formally derived by performing a metric variation of the action presented in equation \ref{1}, which subsequently yields:
 \begin{align}\label{2}
	&\kappa T_{\mu\nu}=-\frac{f}{2}g_{\mu\nu}+\frac{2}{\sqrt{-g}}\partial_{\alpha}\bigg(\sqrt{-g}f_{Q}P^{\alpha}_{\mu\nu}\bigg)\\\nonumber
	&+\bigg(P_{\mu\eta\beta}Q_{\nu}^{\eta\beta}-2P_{\eta\beta\nu}Q^{\eta\beta}{\mu}\bigg)f{Q}\\\nonumber
	&+\bigg(\frac{C}{2}g_{\mu\nu}-\overset{\circ}\nabla_{\mu}\overset{\circ}{\nabla_{\nu}}+g_{\mu\nu}\overset{\circ}{\nabla^{\eta}}\overset{\circ}\nabla_{\eta}-2P^{\alpha}{\mu\nu}\partial{\alpha}\bigg)f_{C},
\end{align}

The $f(Q,C)$ theory of gravity is driven by the need to generalize gravity by offering new ways to explain cosmic acceleration and unify different geometrical frameworks. This motivation aligns with recent research, such as Jimenez et al. \cite{47} discusses the geometric trinity of gravity, where curvature, torsion, and non-metricity offer different perspectives on gravitational theory. It introduces generalizations such as $f(Q,C)$ gravity by considering the non-metricity scalar and boundary terms, Frusciante \cite{48} provides a detailed exploration of the $f(Q,C)$ framework and its cosmological implications. It highlights the role of the boundary term C in modifying gravitational interactions and offers observational signatures that could distinguish $f(Q,C)$ gravity from other models.

Zhao \& Cai \cite{49} delves into the dynamical behavior of $f(Q,C)$ gravity in the context of cosmology. It emphasizes how the boundary term affects the cosmological evolution and the theory's ability to address the accelerating expansion of the universe. Anagnostopoulos {\it et al.} \cite{50}, examines the stability and cosmological consequences of $f(Q,C)$ gravity, focusing on the influence of the boundary term on the evolution of the universe and its potential to explain dark energy. Following the motivational studies on $f(Q,C)$ gravity by multiple authors \cite{50a,50b,50c,50d,50e,50f,50g} , we extend this research by conducting a comprehensive analysis within the same theoretical framework. The outline of the comprehensive analysis is as follows: In section II, we discussed the metric and the field equations, in section III, we present the observational constraints and the results. Some physical parameters of the model is presented in the section IV whereas in section V we provided the concluding remark.

\section{Metric and field equations}\label{1}
This paper considers a homogeneous and isotropic universe, described by the Friedmann-Lemaître-Robertson-Walker (FLRW) spacetime with the following form:
\begin{equation}\label{3}
	ds^{2}=dt^{2}-a^{2}(t) dr^{2}-a^{2}(t)r^{2} d\theta ^{2}-a^{2}(t)r^2\sin^{2}\theta d\phi ^{2},  
\end{equation}
where $a(t)$ represents the universe scalar factor which is dependent on the cosmic time $t$,  and $\left( t,r,\theta,\phi\right)$ denotes the comoving coordinates. 

The stress-energy tensor is provided by the following when we consider the matter to be a perfect fluid:
\begin{equation} \label{4}
	T_{\mu \nu }=\left( \rho +p\right) u_{\mu }u_{\nu }-pg_{\mu \nu },  
\end{equation}
where the four-velocity is $u_{\mu }$ which follows $u_{\mu }u^{\mu }=1$, the  energy density is $\rho $ and isotropic pressure is $p$.\\
By employing equations (\ref{3}) and (\ref{5}), we formally derive the field equations, which take the form:
\begin{equation} \label{5}
\rho=6H^2f_Q-(9H^2+3\dot{H})f_C+3H\dot{f}_C+f/2 
\end{equation}

\begin{equation} \label{6}
	p=-(6H^2+2\dot{H})f_Q-2H\dot{f}_Q+(9H^2+3\dot{H})f_C-\ddot{f}_C-f/2 
\end{equation}
 The field equations (\ref{5}) and (\ref{6}) exhibit nonlinear behavior, rendering their solutions challenging to obtain. To address this, we explore a nonlinear $f(Q, C)$ gravity model of the form \cite{51}:
\begin{equation} \label{7}
	f(Q,C)=a_1 Q^\alpha +a_2 C
\end{equation}
The motivation for choosing the $f(Q,C)=a_1 Q^\alpha +a_2 C$ model lies in its potential to extend General Relativity by incorporating non-metricity and boundary terms. This approach provides a flexible framework for addressing the late-time accelerated expansion of the universe without relying solely on dark energy or a cosmological constant. By introducing the non-metricity scalar $Q$ and the boundary term $C$ the model gains additional degrees of freedom, allowing it to offer a geometric explanation for cosmic acceleration and match a variety of cosmological observations. The power-law dependence on $Q$ and the inclusion of $C$ make the model adaptable for fitting observational data and provide testable deviations from the standard $\Lambda$CDM model, while remaining consistent with current and future experiments \cite{52,53,54,55}.
By incorporating the model described in equation (\ref{7}), the field equations formulated in equations (\ref{5}) and (\ref{6}) can be re-expressed as:
\begin{equation} \label{8}
\rho = -a_1 2^{\alpha -1} 3^{\alpha } (2 \alpha -1) (- H^2)^{\alpha}
\end{equation}

\begin{equation} \label{9}
p = 6^{\alpha -1}a_1 (2 \alpha -1) (-H)^{\alpha } H^{\alpha -2} \left(3 H^2+2 \alpha  \dot{H}\right)
\end{equation}
Subsequently, we explored additional physical parameters that are intimately linked to the energy density and isotropic pressure of the universe. These parameters include the equation of state parameter, the $\left(\omega-\omega^{\prime}\right)$- plane, and energy conditions. A thorough examination of these parameters is essential for elucidating the physical interpretation of the universe.\\
\textit{The equation of state parameter}\\
The equation of state parameter ($\omega$), defined as $\omega=p/\rho$. The equation of state parameter distinguishes various Dark Energy (DE) models. Astrophysical observations suggest that this parameter is approximately -1, indicating a constant energy density with negative pressure, characteristic of the cosmological constant ($\omega = -1$). Beyond the  $\omega = -1$, dynamical DE models can be categorized into: Quintessence models: $\omega > -1$; Phantom models: $\omega < -1$; K-essence models: dynamic $\omega$; Brane cosmology models: variable $\omega$. These categories reflect distinct evolutionary trajectories for the universe, differing from the static $\Lambda$CDM scenario. Hence from above equations (\ref{8}) and (\ref{9}), the $\omega$ is observed as

\begin{equation} \label{10}
\omega=-\frac{1}{3} (-H)^{\alpha } H^{\alpha -2} \left(-H^2\right)^{-\alpha } \left(3 H^2+2 \alpha  \dot{H}\right)
\end{equation}
\textit{The $\left(\omega-\omega^{\prime}\right)$- plane}\\
The $(\omega-\omega')$-plane is a valuable tool for understanding the dynamics of dark energy and its impact on the evolution of the universe. Here, $\omega$ represents the equation of state (EoS) parameter, while $\omega'$ denotes the derivative of $\omega$ with respect to the natural logarithm of the scale factor. Mathematically it is observed as
\begin{equation}\label{11}
	\omega'=\frac{\partial \omega}{\partial (lna)}
\end{equation}
Here, the value of $\omega$ plays a crucial role in determining the fate of the universe. A value of $\omega$ less than $-1/3$ indicates a universe that will continue to expand indefinitely, while a value greater than $-1/3$ suggests a universe that will eventually collapse. The derivative $\omega'$ provides additional information about the evolution of dark energy over time. In the context of the $(\omega-\omega')$-plane, the evolution of the universe can be broadly categorized into four distinct phases: quintessence-like behavior, phantom-like behavior, cosmological constant-like behavior, and transient behavior.\\

\textit{The stability of the model}\\
The stability of a cosmological model is often assessed by examining the squared velocity of sound, denoted as $\vartheta_s^2$. This parameter is a crucial diagnostic tool for understanding the evolution of the universe. Mathematically, $\vartheta_s^2$ is defined as:
\begin{equation}\label{11a}
\vartheta_s^2 = \frac{\partial p}{\partial \rho},
\end{equation}
where $p$ is the pressure and $\rho$ is the energy density.

A positive value of $\vartheta_s^2$ indicates stability, implying that the universe will continue to evolve in a predictable manner. Conversely, a negative value of $\vartheta_s^2$ signals instability, potentially leading to the formation of singularities or other catastrophic events.\\

\textit{The energy conditions}\\
The energy conditions are fundamental constraints in general relativity, ensuring that the energy-momentum tensor of matter and energy satisfies certain physical requirements. These conditions play a crucial role in determining the evolution of the universe, particularly in the context of dark energy and modified gravity theories.\\	
The null energy condition (NEC) states that for any null vector $k^\mu$, the energy-momentum tensor $T_{\mu\nu}$ satisfies the inequality: $T_{\mu\nu}k^\mu k^\nu \geq 0$. The dominant energy condition (DEC) is a stronger requirement than the NEC. It states that for any future-directed timelike vector $u^\mu$, the energy-momentum tensor satisfies: $T_{\mu\nu}u^\mu u^\nu \geq 0 \quad \text{and} \quad T_{\mu\nu}u^\mu$ is future-directed and the strong energy condition (SEC) is the strongest of the three energy conditions. It states that for any future-directed time-like vector $u^\mu$, the energy-momentum tensor satisfies: $(T_{\mu\nu} - \frac{1}{2}Tg_{\mu\nu})u^\mu u^\nu \geq 0$.\\
The values of these energy conditions affect the evolution of the universe in several ways. For instance, the NEC, DEC, and SEC influence the cosmological expansion, dark energy, black hole formation, and cosmological singularities. Mathematically, the energy conditions can be expressed in terms of energy density ($\rho$) and isotropic pressure ($p$) as follows:
 \begin{equation}\label{12}
	\begin{split}
		& NEC: \rho + p \geq 0\\
		& DEC: \rho \geq 0 \;\; \text{and}\;\; \rho + p \geq 0 \\
		& SEC: \rho + 3p \geq 0  \\
	\end{split}
\end{equation}

Next, to facilitate the solution of the field equations and elucidate the physical behavior of the universe, explicit solutions necessitate additional assumptions. In the subsequent section, we employ a straightforward and simple parametrization method for the Hubble parameter ($H$). This approach is crucial as it enables us to, Analytically solve the field equations, providing valuable insights into the cosmological evolution, Investigate the physical implications of model and Examine the viability of various dark energy scenarios. By adopting this parametrization method, we can systematically explore the cosmological implications of different models, thereby gaining a deeper understanding of the universe's evolution.

\begin{equation}\label{13}
	H(z)=b_0\left(b_1+(1+z)^{b_2}\right)
\end{equation}
where $b_0=\frac{H_0}{1+b_1}$, \(b_1\), and \(b_2\) are the model parameters and \(H_0\) denotes the present-day Hubble constant. This parametric form, which corresponds to the so-called transit scale factor $\left(H(z) = \epsilon \left(R^{-b_1} + b_2 \right), \epsilon \;\text{be any arbitrary parameters}\right)$, is particularly useful in describinga smooth transition between different cosmic epochs, accommodating deceleration at early times and acceleration in the present era. Using observational datasets, the model parameters \(H_0\), \(b_1\), and \(b_2\) are constrained using the Markov Chain Monte Carlo (MCMC) method.

\section{Observational constraints and Results}
	\subsection*{A. Hubble Datasets}
	The Hubble parameter, \(H(z)\), is directly related to the redshift (\(z\)) through the differential relationship \(\frac{dz}{dt}\), allowing its determination via spectroscopic surveys. For this analysis, we employ 57 observational data points for \(H(z)\) derived using the differential age method. The theoretical and observed values of \(H(z)\) are compared through a chi-square function:
\begin{equation}\label{14}
	\chi^2_{H(z)} = \sum_{i=1}^{57} \frac{\left[H_{\text{th}}(z_i) - H_{\text{obs}}(z_i)\right]^2}{\sigma^2_{H(z_i)}},
\end{equation}
where \(H_{\text{th}}\) and \(H_{\text{obs}}\) represent the theoretical and observed values of \(H(z)\), respectively, and \(\sigma_{H(z_i)}\) is the observational error.
	
	\subsection*{B. Standard Candles (SC)}
	We incorporate data from the Pantheon Type Ia supernovae, quasars, and gamma-ray bursts (GRBs) to constrain the model. The observed distance modulus \(\mu_{\text{obs}}\) is compared to the theoretical value \(\mu_{\text{th}}\), expressed as:
	\begin{equation}\label{15}
		\mu = m - M = 5 \log_{10}(D_L) + \mu_0,
	\end{equation}
where \(m\) and \(M\) are the apparent and absolute magnitudes, respectively, and \(\mu_0 = 5 \log_{10}\left(H_0^{-1}/\text{Mpc}\right) + 25\). The luminosity distance is given by:
	\begin{equation}\label{16}
		D_L(z) = \frac{c}{H_0} (1+z) \int_0^z \frac{dz^*}{E(z^*)},
\end{equation}
	where \(E(z^*) = \frac{H(z^*)}{H_0}\). The chi-square function for the supernova data is:
		\begin{equation}\label{17}
		\chi^2_{\text{SN}} = \mu_s C_{\text{s,cov}}^{-1} \mu_s^T,
	\end{equation}
where \(C_{\text{s,cov}}\) represents the covariance matrix for the supernova data.
	
	\subsection*{C. Uncorrelated Baryon Acoustic Oscillations (unCor BAO)}
	For BAO data, we use a subset of 17 uncorrelated measurements out of a larger dataset of 333 to reduce errors caused by correlations. The angular diameter distance, \(D_A\), and the Hubble distance, \(D_H\), are expressed as:
		\begin{equation}\label{18}
		D_H(z) = \frac{c}{H(z)}, \quad D_M = (1+z) D_A = \frac{c}{H_0} S_k\left(\int_0^z \frac{dz'}{E(z')}\right),
	\end{equation}
	where \(S_k(x)\) depends on the curvature parameter \(\Omega_k\). The chi-square for the BAO dataset is incorporated into the total likelihood analysis.
	
	\subsection*{D. Monte Carlo Markov Chain (MCMC)}
	The MCMC method is used to constrain the model parameters by minimizing the total chi-square function:
		\begin{equation}\label{19}
		\chi^2 = \chi^2_{H(z)} + \chi^2_{\text{SN}} + \chi^2_{\text{unCorBAO}}.
	\end{equation}
This technique ensures that the model parameters are statistically consistent with the observational data, providing best-fit values and confidence intervals.
	
	\subsection*{E. Results}
\begin{figure}[H]
		\centering
		\includegraphics[scale=0.55]{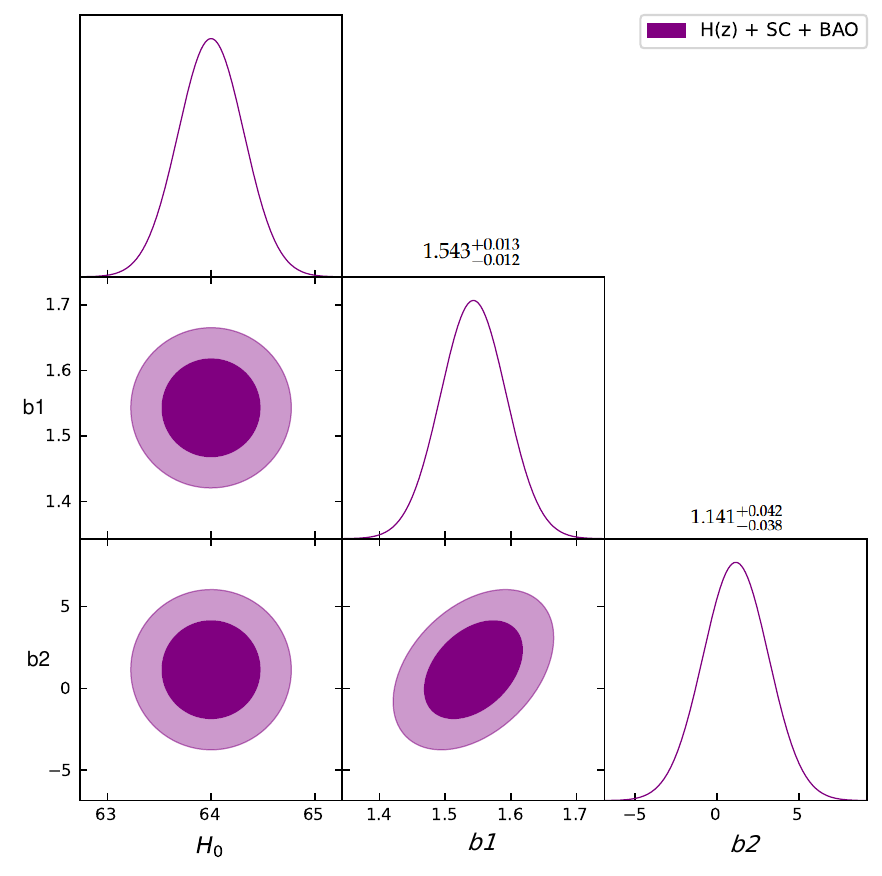}
		\caption{This figure corresponds to $ 1\sigma $ and $ 2\sigma $ confidence contours obtained from  $H(z)+ SC + unCor BAO$ dataset showing $H_{0}, b_1, b_2$ obtained for the transit form of Hubble parameter model.}
		\label{LCDM}
	\end{figure}

	Using the MCMC method, the best-fit values for the model parameters $H_0=64.51^{+0.21}_{-0.20}$, $b_1=1.543^{+0.013}_{-0.012}$, $b_2=1.141^{+0.042}_{-0.038}$ are determined (See Fig. \ref{LCDM}). The parameter constraints indicate a transition redshift that aligns well with current observational data. Figure 1 illustrates the 1\(\sigma\) and 2\(\sigma\) confidence contours for the parameters, demonstrating the robustness of the fit.

\section{Physical parameters of the model}
\subsubsection{The energy density}
\begin{figure}[H]
	\centering
	\includegraphics[scale=0.7]{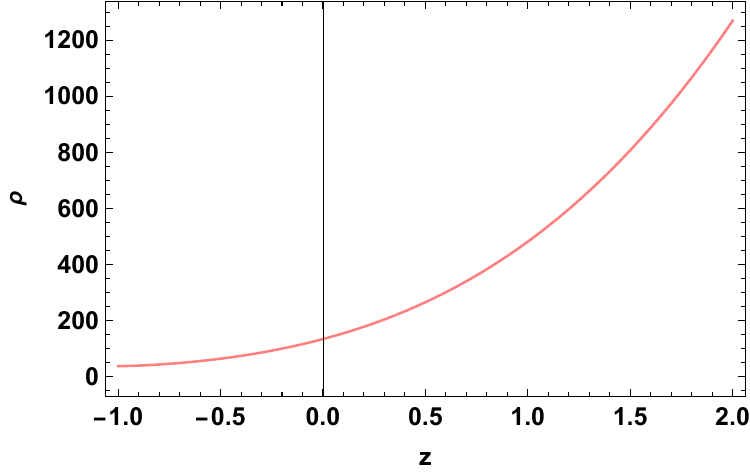}
	\caption{The behavior of energy density for the transit form of Hubble parameter model.}
	\label{den}
\end{figure}

The energy density is mathematically expressed as given in Equation (\ref{8}). To visualize its graphical behavior, we substitute Equation (\ref{13}) into Equation (\ref{8}). The resulting graphical representation of the energy density is illustrated in Figure \ref{den}. The figure shows the evolution of the energy density \(\rho(z)\) in \(f(Q,C)\) gravity, derived using a transitional scale factor. In the early universe (\(z > 1\)), the energy density is significantly high, reflecting the dominance of matter and radiation in a highly energetic state. As redshift decreases towards \(z = 0\), representing the present epoch, \(\rho(z)\) declines steadily, signifying the transition from a matter-dominated phase to one dominated by dark energy. This decrease aligns with the observed expansion of the universe, where the influence of dark energy becomes more pronounced, driving cosmic acceleration. In the future (\(z < 0\)), \(\rho(z)\) asymptotically approaches near-zero values, indicating a transition to a vacuum-dominated de Sitter phase. 
This behavior highlights the ability of \(f(Q,C)\) gravity to model the universe's evolution across its critical phases: early high-energy dominance, present accelerated expansion, and eventual vacuum domination. The model aligns well with the standard \(\Lambda\)CDM framework while incorporating dynamic features of the non-metricity scalar \(Q\) and boundary term \(C\), offering a robust description of cosmic evolution.

\subsubsection{The isotropic pressure}
\begin{figure}[H]
	\centering
	\includegraphics[scale=0.7]{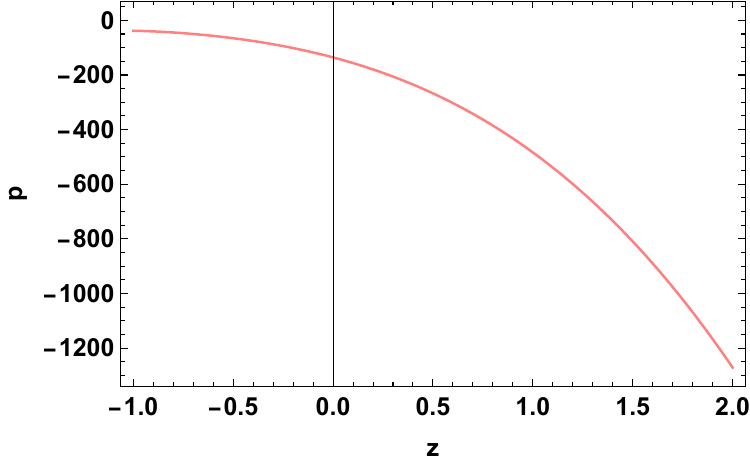}
	\caption{The behavior of isotropic pressure for the transit form of Hubble parameter model.}
	\label{p}
\end{figure}

The isotropic pressure is mathematically expressed as given in Equation (\ref{9}). To visualize its graphical behavior, we substitute Equation (\ref{13}) into Equation (\ref{9}). The resulting graphical representation of the equation of state parameter is illustrated in Figure \ref{p}.
The figure \ref{p} demonstrates the evolution of isotropic pressure (\(p\)) in \(f(Q,C)\) gravity, derived from the Hubble parameter using a transitional scale factor. The curve begins at highly negative values in the early universe (\(z > 1\)) and increases (remaining negative) as \(z\) approaches \(-1\), where it asymptotically converges to zero. This behavior encapsulates the dynamic evolution of the universe across different cosmic epochs, showcasing the capacity of \(f(Q,C)\) gravity to describe transitions between phases. \\
As the universe evolves into the present epoch (\(z \approx 0\)), the pressure increases negatively, reflecting the influence of dark energy. The negative isotropic pressure becomes the primary driver of accelerated expansion, as confirmed by observations such as Type Ia supernovae, CMB, and BAO measurements. The behavior of pressure during this phase aligns well with the requirements for a dark energy-dominated universe, ensuring consistency with observationally supported acceleration. In the far future (\(z < 0\)), the pressure approaches zero, indicating a vacuum-dominated de Sitter-like phase characterized by perpetual accelerated expansion and aligns with a cosmological constant-like scenario in the universe's late-time dynamics. This behavior reflects the ability of \(f(Q,C)\) gravity to unify the early, present, and late phases of the universe while accommodating dynamic deviations from the standard \(\Lambda\)CDM model. The model's flexibility stems from the interplay between the non-metricity scalar \(Q\) and the boundary term \(C\), which effectively govern the evolution of dark energy.

\subsubsection{The equation of state parameter}

\begin{figure}[H]
	\centering
	\includegraphics[scale=0.65]{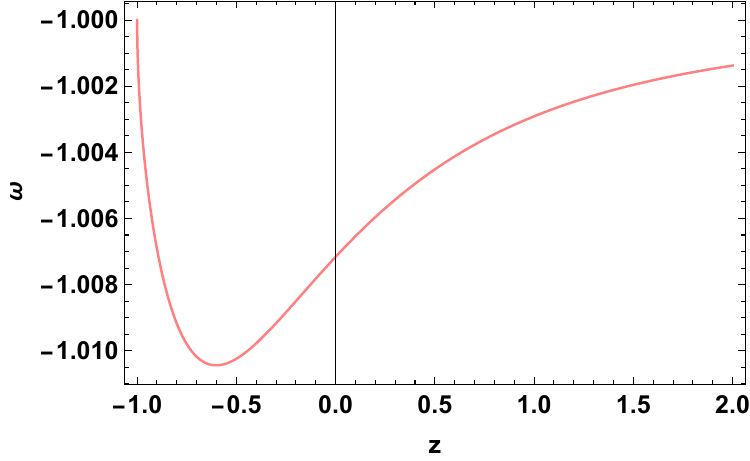}
	\caption{The behavior of equation of state parameter for the transit form of Hubble parameter model.}
	\label{eos}
\end{figure}

The equation of state parameter is mathematically expressed as given in Equation (\ref{10}). To visualize its graphical behavior, we substitute Equation (\ref{13}) into Equation (\ref{10}). The resulting graphical representation of the equation of state parameter is illustrated in Figure \ref{eos}.
The evolution of the equation of state (EoS) parameter \(\omega(z)\) within \(f(Q,C)\) gravity, derived from the Hubble parameter based on a transitional scale factor, reflects the dynamics of the universe's expansion from the early epochs through the present era and into the far future. In the early universe (\(z > 1\)), the EoS parameter is approximately \(-1\), suggesting an early quasi-de Sitter phase where the influence of dark energy is present but not dominant and the dark energy contributions remain minimal yet significant enough to maintain a near-constant EoS. This subtle deviation from \(-1\) reflects the dynamism inherent in the \(f(Q,C)\) framework, highlighting its ability to model evolving dark energy components influenced by the interaction of the non-metricity scalar \(Q\) and boundary term \(C\). As the universe transitions towards the present epoch (\(z \approx 0\)), \(\omega(z)\) stabilizes near \(-1\), aligning closely with observational data from Type Ia supernovae, CMB, and BAO measurements. This near-constant value supports the interpretation of dark energy as the primary driver of the accelerating expansion observed today. The slight deviation from exactly \(-1\) emphasizes the flexibility of \(f(Q,C)\) gravity. In the distant future (\(z < 0\)), \(\omega(z)\) asymptotically approaches exactly \(-1\), indicating a de Sitter phase where dark energy fully dominates the dynamics of the cosmos. This convergence suggests a steady state of perpetual accelerated expansion, consistent with the predictions of the \(\Lambda\)CDM model but achieved within the broader and more dynamic context of \(f(Q,C)\) gravity. \\
The near-identical behavior of \(\omega(z)\) to \(-1\) across all epochs suggests that \(f(Q,C)\) gravity retains the essential features of the \(\Lambda\)CDM model, making it an attractive candidate for exploring deviations from standard cosmology while remaining consistent with current observational constraints. This makes the model particularly valuable in scenarios requiring a dynamic and adaptable approach to understanding the evolving universe.

\subsubsection{The $\left(\omega-\omega^{\prime}\right)$ plane}
\begin{figure}[H]
	\centering
	\includegraphics[scale=0.65]{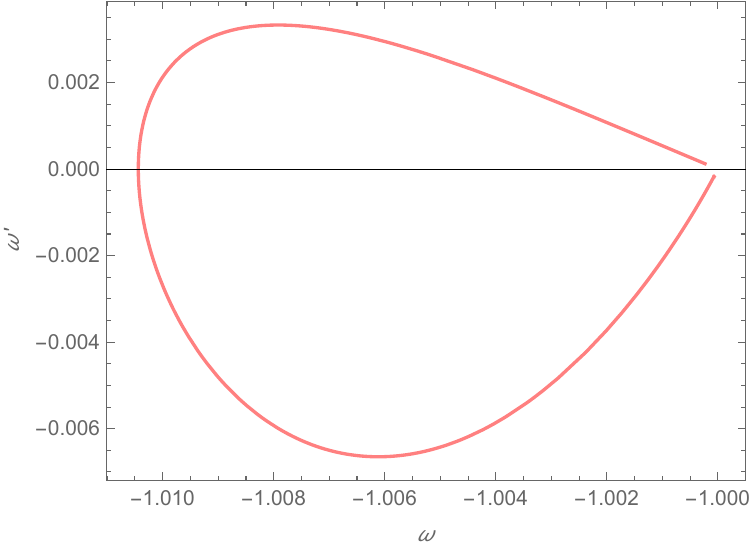}
	\caption{The behavior of squared velocity of sound for the transit form of Hubble parameter model.}
	\label{ww}
\end{figure}

The $\omega^{\prime}$ is mathematically expressed as given in Equation (\ref{11}). To visualize its graphical behavior, we substitute Equation (\ref{13}) into Equation (\ref{11}). The resulting graphical representation of the equation of state parameter is illustrated in Figure \ref{ww}.

	The \((\omega - \omega')\) plane serves as a diagnostic tool to study the dynamical evolution of the equation of state (EoS) parameter, \(\omega\), and its derivative, \(\omega' = \frac{d\omega}{d\ln a}\), within the \(f(Q,C)\) gravity framework. The trajectory on this plane reveals insights into cosmic evolution across different epochs and the underlying dynamics of dark energy. In the early universe (\(z > 1\)), the trajectory starts in the vicinity of \(\omega \approx -1\) with a small positive \(\omega'\). This behavior indicates that the EoS parameter is nearly constant, consistent with a phase dominated by matter and radiation. The deviation from \(\omega = -1\) reflects the influence of the non-metricity scalar \(Q\) and boundary term \(C\), which contribute to the dynamical evolution characteristic of \(f(Q,C)\) gravity. In the present epoch (\(z \approx 0\)), \(\omega\) stabilizes near \(-1\) while \(\omega'\) approaches zero. This behavior aligns with the dominance of a \(\Lambda\)-like dark energy component driving the accelerated expansion of the universe which consistent with observational constraints, including Type Ia supernovae and baryon acoustic oscillation (BAO) data and in the late universe (\(z < 0\)), the trajectory converges at \((\omega, \omega') = (-1, 0)\), corresponding to a de Sitter phase.
	
\subsubsection{The stability of the model}
\begin{figure}[H]
	\centering
	\includegraphics[scale=0.7]{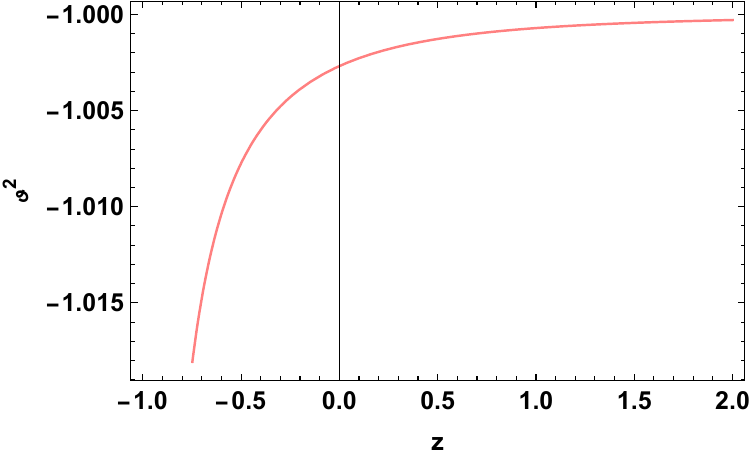}
	\caption{The behavior of squared velocity of sound for the transit form of Hubble parameter model.}
	\label{st}
\end{figure}

The squared velocity of sound is mathematically expressed as given in Equation (\ref{11a}). To visualize its graphical behavior, we substitute Equation (\ref{13}) into Equation (\ref{11a}). The resulting graphical representation of the equation of state parameter is illustrated in Figure \ref{st}.

The figure demonstrates the evolution of the stability parameter (\(c_s^2\)) in \(f(Q,C)\) gravity, derived using a transitional scale factor. The curve reveals a negative \(c_s^2\) throughout, spanning from the early universe (\(z > 0\)) to the far future (\(z < 0\)). In the early universe (\(z > 1\)), the negative values indicate perturbative instabilities arising from rapid density fluctuations.

As the universe transitions into the present epoch (\(z \approx 0\)), the stability parameter remains negative, reflecting challenges in maintaining perturbative stability during dark energy dominance. Despite this, the model remains consistent with observations, including Type Ia supernovae and CMB data, due to its compatibility with large-scale structure formation which underscores the role of the non-metricity scalar \(Q\) and boundary term \(C\) in shaping the dynamics of \(f(Q,C)\) gravity. While the persistent negative values raise concerns about stability.
\subsubsection{The energy conditions}
\begin{figure}[H]
	\centering
	\includegraphics[scale=0.7]{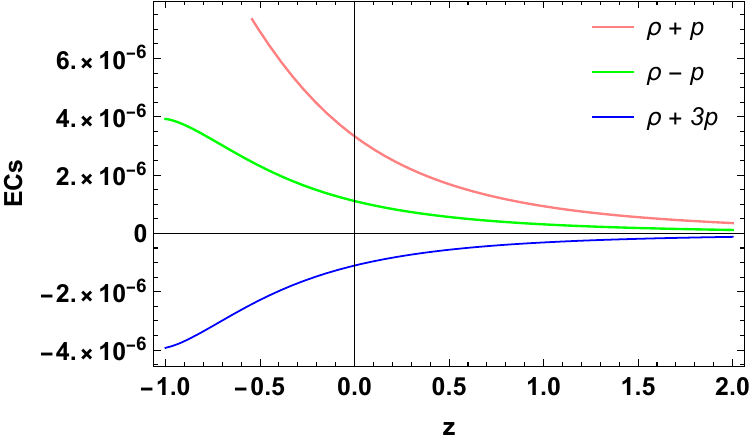}
	\caption{The behavior of energy conditions for the transit form of Hubble parameter model.}
	\label{ec}
\end{figure}

The energy conditions is mathematically expressed as given in Equation (\ref{12}). To visualize its graphical behavior, we substitute Equation (\ref{13}) into Equation (\ref{12}). The resulting graphical representation of the equation of state parameter is illustrated in Figure \ref{ec}.

The figure illustrates the behavior of the Null Energy Condition (NEC), Dominant Energy Condition (DEC), and Strong Energy Condition (SEC) in \(f(Q,C)\) gravity, derived using the Hubble parameter from a transitional scale factor. The NEC and DEC exhibit positively decreasing behavior from \(z > 0\) (early universe) to \(z = -1\) (future universe), while the SEC is consistently violated throughout the same redshift range. This behavior encapsulates the dynamics of the universe’s expansion and the interplay between energy components in \(f(Q,C)\) gravity. 

In the early universe (\(z > 1\)), the NEC and DEC are positive and relatively high, reflecting the dominance of radiation and matter, which satisfy these conditions under the framework of general relativity and modified gravity theories. These positive values ensure the physical viability of \(f(Q,C)\) gravity during the early epochs, as the NEC and DEC are fundamental to ensuring non-negative energy density and the dominance of gravitationally attractive matter. The SEC, however, is violated even in the early universe, highlighting the influence of dark energy-like components in driving accelerated expansion. The SEC violation is consistent with observations of early cosmic acceleration, such as those inferred from the inflationary epoch, where repulsive forces dominate due to negative pressure. In the far future (\(z < 0\)), the NEC and DEC approach near-zero values while remaining positive, signifying the asymptotic approach to a de Sitter-like vacuum-dominated phase. This phase corresponds to perpetual accelerated expansion, where the energy density becomes nearly constant, and the effects of matter and radiation are negligible. The SEC remains violated in this regime, consistent with the dominance of a cosmological constant-like component driving the expansion. The persistent SEC violation underscores the ability of \(f(Q,C)\) gravity to model a universe that transitions smoothly from decelerating expansion (dominated by matter and radiation) to accelerating expansion (dominated by dark energy) and finally to a stable vacuum-dominated state. The role of the non-metricity scalar \(Q\) and boundary term \(C\) in modulating these energy conditions reflects the dynamical nature of \(f(Q,C)\) gravity, making it a robust alternative to standard cosmology.
Observational evidence, including Type Ia supernovae, CMB anisotropies, and BAO data, supports this behavior, validating the \(f(Q,C)\) gravity framework as an extension of the \(\Lambda\)CDM model with dynamic energy conditions.

\section{Conclusion}

The investigation of $f(Q,C)$ gravity using a transition-based Hubble parameter has provided profound insights into the physical behavior of key cosmological parameters across different epochs. The model successfully describes the evolution of the universe, encompassing early deceleration and late-time acceleration, while remaining consistent with observational datasets.

The energy density demonstrates a gradual decrease from high values in the early universe $(z > 0)$ to near-zero values in the far future $(z < 0)$, consistent with the transition from a radiation-dominated epoch to a vacuum-dominated de Sitter-like phase. This behavior underscores the model's ability to accommodate the diminishing influence of matter and radiation as the universe expands. The isotropic pressure evolves from highly negative values in the early universe to values approaching zero in the late universe. This trend reflects the dynamic of accelerated expansion driven by dark energy. The increasing negativity of the pressure highlights the role of $f(Q,C)$ gravity in capturing the effects of repulsive forces required for cosmic acceleration. The equation of state parameter ($\omega$) remains close to -1 in the early universe and stabilizes exactly at -1 in the far future, aligning with observations of a cosmological constant-like behavior. This parameter's evolution signifies the presence of dark energy-like dynamics at early times and its eventual dominance in the universe's late-stage evolution. In our analysis, the \((\omega - \omega')\) plane not only illustrates the dynamical evolution of dark energy but also serves as a key diagnostic to distinguish \(f(Q,C)\) gravity from other cosmological models. The trajectory's alignment with the \(\Lambda\)CDM baseline during the present epoch, combined with its deviations in earlier epochs, highlights the model's capacity to capture unique dynamical features while maintaining consistency with observational data. This analysis validates \(f(Q,C)\) gravity as a promising alternative framework for understanding the universe's evolution.

The analysis of stability parameters reveals a consistently negative behavior, highlighting perturbative instabilities that arise naturally within $f(Q,C)$ gravity. While this poses challenges, it also suggests non-standard mechanisms of cosmic structure formation that distinguish this framework from $\Lambda$CDM. The study of energy conditions further validates the model’s physical viability. The null energy condition (NEC) and dominant energy condition (DEC) remain satisfied across all epochs, ensuring the model's consistency with fundamental physical principles. However, the strong energy condition (SEC) is violated throughout, which is essential for explaining the observed accelerated expansion of the universe.

In summary, the $f(Q,C)$ gravity framework effectively models the universe's evolution, providing a unified explanation of its physical behavior across early, present, and late epochs. By addressing the dynamics of isotropic pressure, energy density, stability parameters, and energy conditions, this model offers a comprehensive alternative to $\Lambda$CDM.

\section*{Declaration of competing interest}
The authors declare that they have no known competing financial interests or personal relationships that could have appeared to influence the work reported in this paper.

\section*{Data availability}
No data was used for the research described in the article.

\section*{Acknowledgments}
The IUCAA, Pune, India, is acknowledged by the authors (A. Pradhan, A. Dixit \& S. H. Shekh) for giving the facility through the Visiting Associateship programmes. Also, this research is funded by the Science Committee of the Ministry of Science and Higher Education of the Republic of Kazakhstan (Grant No. AP23483654).

\end{document}